\documentclass[]{spie}  

\usepackage{amsmath,amsfonts,amssymb}
\usepackage{graphicx}
\usepackage[colorlinks=true, allcolors=blue]{hyperref}
\newcommand{\as}{\prime\prime}

\title{Balloon UV Experiments for Astronomical and Atmospheric Observations}

\author[1]{A.G. Sreejith}
\author[1]{Joice Mathew}
\author[1]{Mayuresh Sarpotdar}
\author[1]{K. Nirmal}
\author[1]{S. Ambily}
\author[1]{Ajin Prakash}
\author[1]{Margarita Safonova}
\author[1]{Jayant Murthy}

\affil[1]{Indian Institute of Astrophysics, Bangalore, India}

\authorinfo{Send correspondence to A.G. Sreejith.\\E-mail: agsreejith@iiap.res.in}

\begin{document} 
\maketitle

\begin{abstract}

The ultraviolet (UV) window has been largely unexplored through balloons for astronomy. We discuss here the development of a compact near-UV spectrograph with fiber optics input for balloon flights. It is a modified Czerny-Turner system built using off-the-shelf components. The system is portable and scalable to different telescopes. The use of reflecting optics reduces the transmission loss in the UV. It employs an image-intensified CMOS sensor, operating in photon counting mode, as the detector of choice. A lightweight pointing system developed for stable pointing to observe astronomical sources is also discussed, together with the methods to improve its accuracy, e.g. using the in-house build star sensor and others. Our primary scientific objectives include the observation of bright Solar System objects such as visible to eye comets, Moon and planets. Studies of planets can give us valuable information about the planetary aurorae, helping to model and compare atmospheres of other planets and the Earth. The other major objective is to look at the diffuse UV atmospheric emission features (airglow lines), and at column densities of trace gases. This UV window includes several lines important to atmospheric chemistry, e.g. SO$_{2}$, O$_{3}$, HCHO, BrO. The spectrograph enables simultaneous measurement of various trace gases, as well as provides better accuracy at higher altitudes compared to electromechanical trace gas measurement sondes. These lines contaminate most astronomical observations but are poorly characterized. Other objectives may include sprites in the atmosphere and meteor flashes from high altitude burn-outs. Our recent experiments and observations with high-altitude balloons are discussed.

\end{abstract}


\section{INTRODUCTION}
\label{sec:intro}  

High-altitude balloon experiments operating in stratosphere provide us with a unique opportunity to observe the sky in mostly any wavelength band with low airmass. As we are operating above 99\% of the atmosphere, these experiments allow diffraction-limited operations without the use of adaptive optics. Balloon payloads also experience lower downwelling radiation by a factor of more than ten as compared to the  ground\cite{1}. Floating zero-pressure balloons provide continuous observations varying from a few hours to days, depending on the launch site. The lower cost and the flexibility in launch timings and flight duration make high-altitude balloon experiments a wonderful test bed for space technologies\cite{2}.

Telescopic platforms at high altitudes have significant advantages over operations from the ground enabling observations at wider wavelength range. The weather conditions at these altitudes are also ideal, where the absence of clouds, water vapour and dust provide virtually ideal atmospheric transmission. The wavefront distortion at heights above 36 km is also very small. Wavelengths below 400 nm are difficult to observe from ground-based observatories, but at balloon floating altitudes observations above 280 nm, as well as between 200 and 220 nm, become possible, the latter due the narrowing of O$_{3}$ and O$_{2}$ absorption bands\cite{1}. Thus, a UV telescope (200--400 nm) located at these altitudes with aperture of just 6-inch in diameter, with sufficient pointing stability/accuracy and a 1K$\times 1$K CCD array, could provide wide-field images with FWHM better than $1^{\as}$,  approaching the diffraction limit\cite{3}; similar to that of space observatories but at a much lower cost. Additionally, space-based observatories have strict constrains because of detector safety reasons. These constrains on Sun/Moon angles limit observations of Solar System objects. 

A stratospheric balloon payload typically has three following main components: a zero-pressure hydrogen/helium balloon of the required volume, a long flight train containing the recovery parachute and flight termination systems, and the gondola consisting of scientific equipment with  communication and associated electronics\cite{4}.

\section{SCIENCE Objectives}

\subsection{Astronomical Observations}

One of our major goals is to observe  astronomical targets of opportunity from balloons. These objects appear with little warning and, as a result, are often only observed in the visible from ground-based observatories. With our quick response time we can have a balloon launch within two weeks, being well placed to obtain UV observations of these transients. A prime candidate for such observations would be comets which suddenly appear, as did comet ISON (C/2012 S1) or Siding Spring (C/2013 A1). These were pristine comets whose evolution as traced by the emission in the molecular bands of different ices should have revealed much about their composition. We have developed  pointing/stabilization systems suitable for balloon flights which will allow us to observe comets and other transients (e.g. supernovae) in bands not observable by any other ground-based system. Other high risk targets could be Mercury/Venus, whose observations are difficult because of the proximity to the Sun.

There are only a handful of UV observations of the Moon (surprisingly for such a prominent object in the sky), primarily because the Moon is so bright that most spacecrafts have an avoidance zone around it. We propose to observe the Moon both spectroscopically and with an imager to track changes in the albedo at different phases, where the phase angle of the Solar illumination changes. There is an increased interest of the Moon and it is important to understand the nature of the lunar surface\cite{5}.

\begin{figure}[b!]
\centering
\includegraphics[scale=0.6]{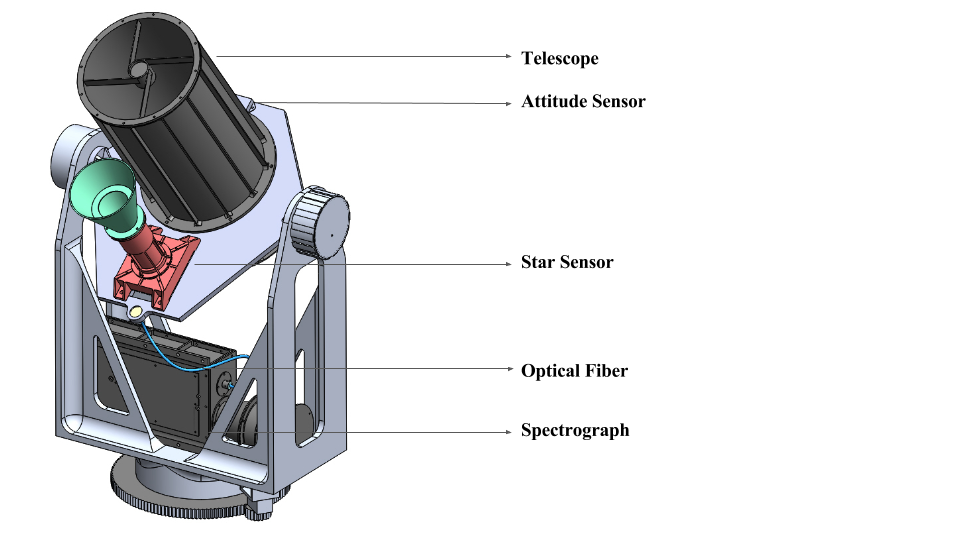}
\caption{\label{fig:payload}The complete payload.}
\end{figure}

\subsection{Atmospheric Observations}

Balloons can provide remote sensing and in-situ measurements of trace gases in upper troposphere and stratosphere\cite{6}. The UV window of 200--400 nm includes the lines from several important species in atmospheric chemistry, such as S0$_{2}$, O$_{3}$, BrO, HCHO. Spectroscopic observations at these wavelengths allow simultaneous measurement of multiple gas species, and thus avoid the need for separate electrochemical systems for each gas. Balloon experiments are complementary to space-based instruments in that they are able to measure gas profiles with different observational geometries, while space-based instruments continuously monitor changes in the atmosphere over long periods of time. The widely accepted DOAS technique can be easily applied to spectroscopic observations of light in the atmosphere (scattered or directly from the source) in UV and visible wavelength regions to obtain the trace gas strengths\cite{7}.

\section{Proposed Payload}
\label{sec:sections}

The overall structural payload design is shown in Fig ~\ref{fig:payload}. The payload will include several scientific instruments for simultaneous observations in both daytime and nighttime. The major instrument for the payload is a spherical catadioptric telescope of 80 mm aperture together with the UV spectrograph. The control of the pointing of the system is achieved through inertial measurement sensors (coarse pointing) and star sensor (fine pointing).   

We will be operating this payload in two different modes. For bright sources, the telescope with a UV-sensitive CCD at the secondary focus will carry out imaging, while the spectrograph, with optical fiber connected to the light-collecting lens (20 mm aperture) aligned with the telescope aperture, will conduct parallel spectroscopic observations of the same bright source. For faint sources observations, the optical fiber from the spectrograph will be placed at the secondary focus of the telescope enabling the spectroscopic observations. Both modes can also be used for atmospheric observations. During day observations, the system will operate with only the coarse pointing as the star sensor will have difficulty identifying stars. 

Individual components of the payload are described in detail below. The float altitudes have extreme environment conditions of low pressure, and the temperature can go down to $-50^{o}$C or so. The low humidity and pressure at these altitudes demand extra care of high-voltage power supplies and detectors to prevent arcing. However, in contrast to space missions, the payload will experience lower vibrations and shocks.    

\subsection{Telescope}

The system consists of an 80 mm spherical catadioptric telescope mounted on the pointing platform\cite{8}. It has a rectangular field of view of $0.46^{\circ}\times 0.34^{\circ}$. The optical design is a two spherical mirror configuration and a double-pass corrector lens system. The secondary focus can house either a UV CCD with a  suitable filter or a slit, depending on the mode of observation. The CCD is a $1360\times1024$ pixels detector which operates at 12 frames per second (fps). The instrument is designed to withstand loads up to 25 g. One of the main objectives is to scan the sky looking for transients. The instrument will acquire the images at a fast frame rate and will analyze each frame looking for brightness variations over the frames. The transient events will be stored on-board or transmitted back to the ground station through a radio link.

\begin{table}[h!]
\caption{Spectrograph Technical Specifications}\label{table:spectrograph}
\centering
\begin{tabular}{lp{10cm}}
\hline
Dimensions & $350\times 130 \times 45$ mm \\
Weight & ~1.5 kg \\
Power & 500 mA at +5 VDC \\
Design & Modified crossed Czerny-Turner \\
Input Fiber & Connector SMA 905 to single-strand optical fiber (0.22 NA) \\
Gratings & Holographic UV, 600 lines/mm \\
Entrance Slit & 100 $\mu$m \\
Sun Avoidance Angle &  Based on the experiment \\
Detector & Micro-Channel Plate MCP40 \\
Wavelength & 250--400 nm \\
\hline
\end{tabular}
\end{table}

\begin{figure}[t!]
\centering
\includegraphics[scale=0.4]{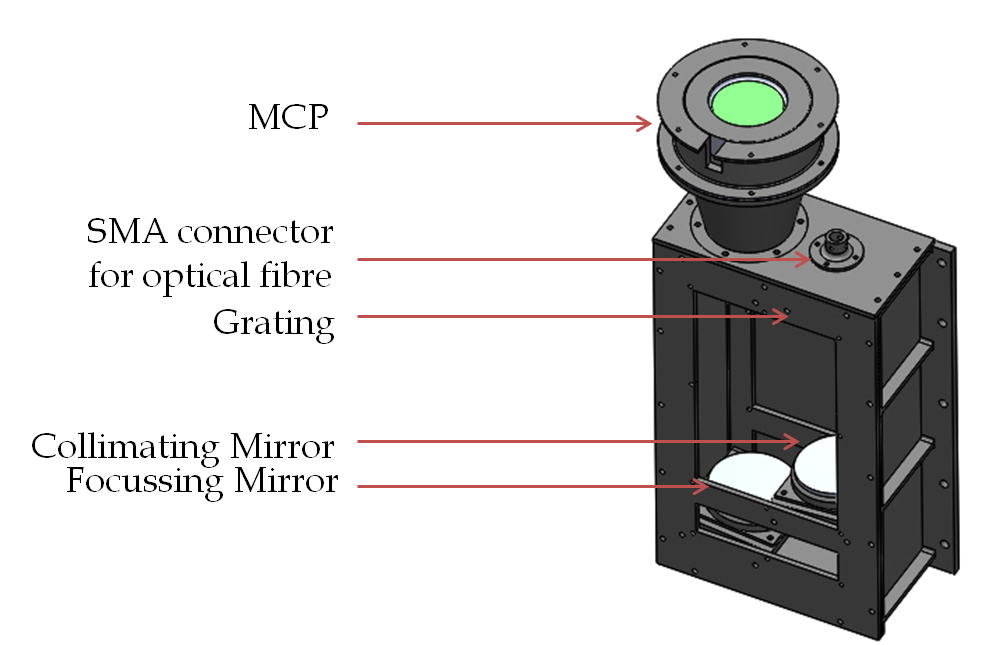}
\caption{\label{fig:spectro}The near ultraviolet spectrograph.}
\end{figure}

\subsection{Spectrograph}

\begin{figure}[b!]
\centering
\includegraphics[width=1\textwidth]{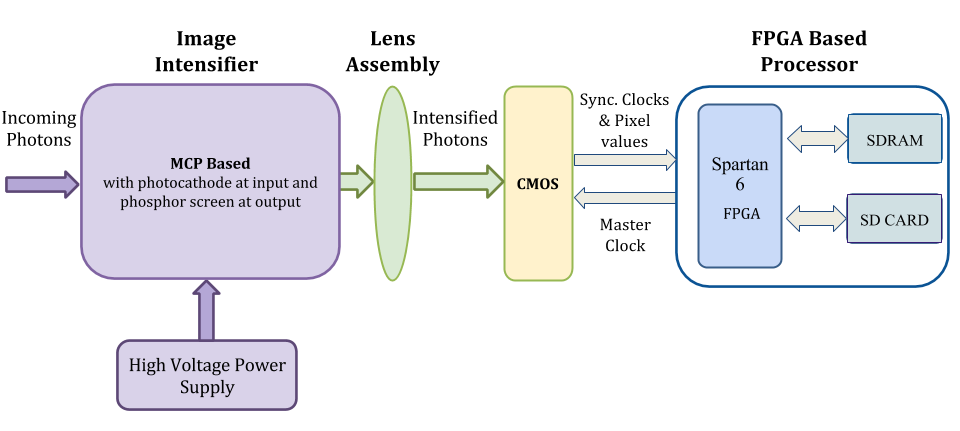}
\caption{\label{fig:detector}The Block diagram of detector.}
\end{figure}

The spectrograph used here is compact fiber-fed modified Czerny-Turner (C-T) design (Fig.~\ref{fig:spectro}).The light is fed to the spectrograph through a $100\,\mu$m dia optical fiber, whose  aperture acts as the entrance slit. The spectrograph is designed to work in the wavelength range of 250--400 nm (for basic design considerations of the spectrograph see Ref.~\citenum{9}). It uses  commercial-off-the-shelf (COTS) optics with an MCP-based image intensified CMOS camera as the detector of choice\cite{ambily}. The detector can operate in either photon counting mode or full frame transfer mode, depending on the mode of operation. The detector readout is controlled by an FPGA board which enables us to do real-time processing and centroiding of the photon events and store the output data on-board. The block diagram of detector system used in the spectrograph is shown in  Fig.~\ref{fig:detector}. The technical details of the spectrograph are given in Table~\ref{table:spectrograph}. The CMOS sensor and the associated optics are attached to the back-end plate of the MCP. The other end of the optical fiber is connected to the telescope, or the light collecting lens, depending on the mode of observation described in the introduction to Section~3.

\subsection{Pointing System}

With a telescope placed on the high-altitude balloon payload, one of the major factor in the design of the pointing system is the transfer of oscillations from the flight train, with this constrain more important if the required stability for observation is within one degree of accuracy. In our first experiments of atmospheric lines (Sreejith et al., 2016), the pointing accuracy was not of a great concern, but we plan to observe astronomical sources for which we require pointing and stabilization of the order of arcseconds.

\begin{figure}[!b]
\centering
\includegraphics[width=.99\textwidth]{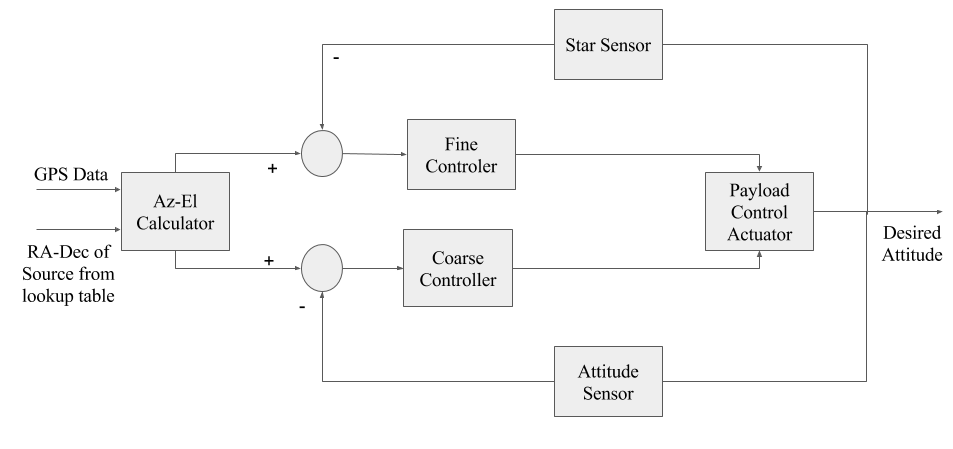}
\caption{\label{fig:block}Block diagram of fine and coarse pointing control.}
\end{figure}

The pointing and stability of the system is carried out as two-fold (coarse and fine pointing) operation, using inertial measurement sensors and a star sensor, respectively. The flow chart describing the pointing system is shown in Fig.~\ref{fig:block}. The attitude sensor obtains the pointing information and corrects the position to an accuracy of $\pm 0.24^{\circ}$ using servomotors. The star sensor works in the inner loop, providing much finer accuracy of around $13^{\as}$.

\subsubsection{Coarse Pointing}

Coarse pointing is carried out using an attitude sensor (Refs.~\citenum{10},~\citenum{11}) which acts as the brain for a closed-loop pointing system using servomotors. This sensor comprises an inertial measurement unit (consisting of 3-axis accelerometer, 3-axis gyroscope and 3-axis magnetometer) with a GPS unit to give pointing to an accuracy of $\pm 0.24^{\circ}$ in either Earth-centered inertial coordinates (azimuth and elevation), or in Right Ascension (RA)  and Declination (Dec). The technical details of the attitude sensor are given in Table~\ref{table:attitude}. This set-up has been tested and found to perform satisfactorily for our requirements\cite{11}.

\begin{table}[h!]
\caption{Attitude Sensor Technical Specifications}\label{table:attitude}
\centering
\begin{tabular}{lp{10cm}}
\hline
Size & $86\times 54\times 45$ mm\\
Weight & $<100$ g without battery\\
Power & 5 W\\
Components & Accelerometer, Gyroscope, Magnetometer and GPS\\
Accuracy & $0.48^{\circ}$ (average RMS)\\
Output Modes & RA-Dec or Az-Ele\\ 
\hline
\end{tabular}
\end{table}

\subsubsection{Fine Pointing}

Fine pointing is achieved through the use of a star sensor, which is a highly sensitive wide-field imaging camera\cite{mico}. The camera has a four-element Tessar lens system as its optical element. Optical design was carried out keeping in mind reduced chromatic aberration, coma and distortion. The star sensor has $10^{\circ}$ field of view  and uses a limiting magnitude of $6.5$. It can provide pointing to an accuracy of $12.24^{\prime\prime}$ for a signal to noise ratio of $30$. The star sensor can work at a maximum slew rate of about 2$^{\circ}$/sec. The technical details of the star sensor are given in Table~\ref{table:star_sensor}. The coarse pointing maintains the payload in this slew rate, and the fine pointing is achieved by the inputs from the star sensor. 

\begin{table}[h!]
\caption{Star Sensor Technical Specifications}\label{table:star_sensor}
\centering
\begin{tabular}{lp{10cm}}
\hline
Accuracy & $12.24^{\prime\prime}$\\
Update Rate & 10 Hz \\
Weight & 800 gms \\
Dimension & $10\times 10\times 50$ cm \\
Power & 2W \\
Communication interface & RS-422 \\
Memory & 8MB Flash and 64 MB RAM \\
Wavelength range & 450--750 nm\\
Sun Avoidance angle & $45^{\circ}$ \\
\hline
\end{tabular}
\end{table}

\section{Flight Opportunity}

We plan to fly the payload on a zero-pressure balloon capable of floating at $\sim 40$ km altitude for up to 5 hours with the National Balloon Facility station of TIFR at Hyderabad\cite{manchanda} in November of 2016. The duration of the float will enable us to observe in the night (with the launch 3--4 hours before sunrise) for astronomical objects, and at daylight for atmospheric studies. The reference spectrum for trace gas analysis will be obtained at float, and the measurements will continue during descent\cite{6}. 

In this flight we also plan to supplement the observational instruments with the detachable pre-sterilized sampling chamber {\it SAMPLE} (Stratospheric Altitude Microbiology Probe for Life Existence)\cite{ajin} designed to collect and contain the dust stratospheric samples and get them back without contamination. November is the peak time for the Leonid meteor shower, and we plan to collect cometary dust particles, where further study will be conducted to establish the possibility of microbial life in the upper atmosphere.

\section{Summary and Conclusion}

We have presented an overview of our ballooning program and the instruments that currently are developed for astronomical observations. These instruments can also be used for atmospheric observations such as measuring the limb scattered UV radiance and trace gas strengths.  

In our earlier high-altitude flights, the payloads experienced frequent oscillations preventing us from continuous observation for a long duration. We have developed a pointing and stabilization platform with two-fold coarse/fine pointing, which will enable stable observations of regions of interest. We are testing and verifying the operation of our payloads on balloon flights with the intention of placing them on small satellites, such as, e.g. CubeSats. Going into space has the obvious advantages of longer mission life and the ability to go further into the far-UV. There is an increasing number of opportunities for small missions although without the quick response times possible with balloons, and the development of instrumentation skills will benefit our future access to space.

\acknowledgements
Part of this research has been supported by the Department of Science and Technology (Government of India) under Grant IR/S2/PU-006/2012.

\end{document}